\documentclass{article}
\usepackage{spconf,amsmath,graphicx}
\usepackage[dvipsnames]{xcolor}
\usepackage{enumitem}
\setlist{nosep, leftmargin=14pt}

\usepackage{algorithm}
\usepackage[noend]{algpseudocode}
\usepackage{amsfonts}
\usepackage{mwe} % to get dummy images
\usepackage{hyperref}

\usepackage{pgfplots}
\usepgfplotslibrary{groupplots}

\usepackage{graphicx}
\usetikzlibrary{positioning}
\usepackage[absolute,overlay]{textpos}

\colorlet{GREEN}{green}

\usepackage{subfig}

\usepackage{caption}
\graphicspath{{img/}}

\def\nblobs{{q}}

\def\realimage{{I}}
\def\realmask{{M}}
\def\realdsindex{{j}}
\def\realdsIndex{{{J}}}
\def\realds{{\{(\realimage_{\realdsindex}, \realmask_{\realdsindex})\}_{\realdsIndex}}}

\def\realimageset{{\{\realimage_{\realdsindex}\}_{\realdsIndex}}}

\def\refstyle{{R}}

\def\nrows{{h}}
\def\ncols{{w}}

\def\blobarea{{A_B}}

\def\realblobsindex{{k}}
\def\realblobs{{B}}
\def\realblobsIndex{{K}}
\def\realblobsset{{\{ \realblobs_{\realblobsindex} \}_{\realblobsIndex}}}

\def\genblobs{{\widetilde{\realblobs}}}
\def\genblobsindex{{l}}

\def\genblobsIndex{{L}}
\def\genblobsset{{\{\genblobs_{\genblobsindex} \}_{\genblobsIndex}}}

\def\genindex{{n}}
\def\genIndex{{N}}

\def\perimeterpoints{{E}}

\def\genmask{{\widetilde{\realmask}}}
\def\genmaskset{{\{\genmask_{\genindex}\}_{\genIndex}}}

\def\genimage{{\widetilde{\realimage}}}
\def\genimageset{{\{\genimage_{\genindex}\}_{\genIndex}}}

\def\gends{{\{(\genimage_{\genindex}, \genmask_{\genindex})\}_\genIndex}}

\def\predictedmask{{\widehat{\realmask}}}

\def\prior{{\mathcal{P}}}
\def\priorsset{{\{\prior_{\genindex}\}_\genIndex}}
\def\availablemask{{A}}
\def\guidingmap{{G}}

\def\priorspacing{{\mathcal{S}}}

\def\trainingset{{\widetilde{D}}}

\def\blobone{{\realblobs_{\realblobsindex1}}}
\def\blobtwo{{\realblobs_{\realblobsindex2}}}

\title{An expert-driven data generation pipeline for histological images}

\copyrightnotice{ \footnotesize \textcopyright 2024 IEEE. Personal use of this material is permitted.
	Permission from IEEE must be obtained for all other uses, in any current or future
	media, including reprinting/republishing this material for advertising or promotional
	purposes, creating new collective works, for resale or redistribution to servers or
	lists, or reuse of any copyrighted component of this work in other works.
	DOI: \href{<http://tex.stackexchange.com>}{<DOI No.>}}
	
\name{Roberto Basla, Loris Giulivi, Luca Magri, Giacomo Boracchi	
	\thanks{Code available at \href{https://github.com/rb-sl/ExpertDrivenNuclei}{https://github.com/rb-sl/ExpertDrivenNuclei}.}}
\address{ \{name.surname\}@polimi.it\,\,\,\,\,\,\,	DEIB, Politecnico di Milano, Italy}

\usepackage{fancyhdr}
\fancypagestyle{firststyle}
{
	\fancyfoot[C]{\scriptsize © 2024 IEEE.  Personal use of this material is permitted.  Permission from IEEE must be obtained for all other uses, in any current or future media, including reprinting/republishing this material for advertising or promotional purposes, creating new collective works, for resale or redistribution to servers or lists, or reuse of any copyrighted component of this work in other works.}
}

\begin{document}
	\maketitle
	
	\thispagestyle{firststyle} 
	\begin{abstract}
		Deep Learning (DL) models have been successfully applied to many applications including biomedical cell segmentation and classification in histological images. These models require large amounts of annotated data which might not always be available, especially in the medical field where annotations are scarce and expensive. 
        To overcome this limitation, we propose a novel pipeline for generating synthetic datasets for cell segmentation. Given only a handful of annotated images, our method generates a large dataset of images which can be used to effectively train DL instance segmentation models. Our solution is designed to generate cells of realistic shapes and placement
        % be modular and tunable, 
        by allowing experts to incorporate domain knowledge during the generation of the dataset.
	\end{abstract}
    \section{Introduction}
\begin{figure*}[t]
		\centering
		\includegraphics[width=\textwidth]{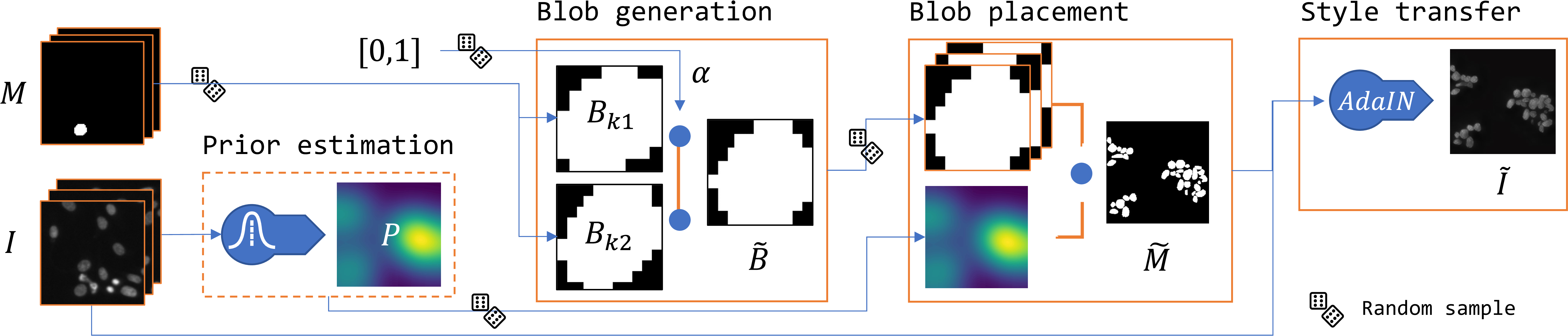}
            \put(-348,101){\fontsize{12}{15}\selectfont {\textcircled{\raisebox{0.4pt} {\fontsize{10}{11}\selectfont $a$}}}}
            \put(-228,101){\fontsize{12}{15}\selectfont {\textcircled{\raisebox{-0.4pt} {\fontsize{10}{11}\selectfont $b$}}}}
            \put(-102,101){\fontsize{12}{15}\selectfont {\textcircled{\raisebox{0.4pt} {\fontsize{10}{11}\selectfont $c$}}}}
  
		\caption{Our generation pipeline. Our first phase is \emph{Blob Generation} \textcircled{\raisebox{0.4pt} {$a$}}, which creates a set of new blobs $\genblobsset$ by interpolating existing ones. We then perform \emph{Blob Placement} \textcircled{\raisebox{-1pt} {$b$}} to generate the GT $\genmask$ following a prior distribution $\prior$ estimated from the few annotated images. Finally, the \emph{Image Generation} \textcircled{\raisebox{0.4pt} {$c$}} phase performs style transfer to transform $\genmask$ into the new image $\genimage$.}
		\label{fig:pipeline}
	\end{figure*}

    In the medical domain, the effectiveness of Deep Learning (DL) methods can be hindered by data scarcity since annotated data may be hard or expensive to obtain.
    This issue is also accentuated by the variety of scenarios DL models have to face in medical imaging, as specific datasets need to be prepared to train DL models on new tissues, when image acquisition techniques change, or to address different goals.
    In this work, we propose a novel pipeline (Fig.~\ref{fig:pipeline}) that enables training instance segmentation models in the histopathological imaging domain in very low data regimes. 
    
	% Common solutions
    Previous works have addressed the issue of data scarcity by means of Data Augmentation (DA) techniques \cite{shorten_survey_2019}. The idea behind DA is to manipulate images in a realistic manner to increase the amount of annotated data during training, thus improving performance.
    % Why difficult - part 2
    These techniques have been extensively used for training classifiers, where supervision is at the image level, as far as they do not change the image's label. For tasks like histological instance segmentation, where annotations are provided at pixel level, the very same transformations often need to be applied both to the image and to the pixel-wise Ground Truth (GT). 
    Common DA strategies include geometric and photometric transformations such as rotations, crops and contrast variations. Unfortunately, they can only be applied to already-existing samples resulting in a limited increase of variability. Image generation, instead, has the potential to obtain a large amount of diverse data, enabling more effective model training. On the flip side, this also requires generating annotations (here also referred to as \textit{blobs}) that are pixel-wise consistent with generated samples.
 
    A few efforts \cite{mahmood_deep_2018, li_high_2022} have been made towards generating both image and GT. These rely on DL models like Generative Adversarial Networks (GANs) \cite{goodfellow_generative_2014} that, while providing good results, do not enable to steer the image generation towards images featuring desired properties like the cell distribution and spacing.
    Other works break down the generation problem to make it more controllable, but are limited to re-using cell masks from real data \cite{mahmood_deep_2018}, or generating blobs at random \cite{hou_robust_2019}, yielding potentially unrealistic results. Other approaches extract blobs from real images and place them over an empty canvas to create the image mask \cite{cheng_deep_2021, kromp_evaluation_2021}. Lastly, works such as \cite{hou_unsupervised_2017} perform style transfer to transform a generated GT into a realistic image in a fully-DL framework. 
 
    None of these methods guarantee that the generated blobs and their positions are coherent with the target tissue, in particular in low-data regimes. We overcome this limitation by proposing a modular pipeline, illustrated in Fig.~\ref{fig:pipeline}, that can incorporate domain knowledge into the blob generation and placement processes. Our pipeline also allows experts to steer them towards realistic samples even at very low data regimes. Our method is composed of three major steps: 
    \textcircled{\raisebox{-1pt} {$a$}}
    Blob Generation, where pairs or randomly selected blobs -- representing different sections of 3D cells -- are interpolated to generate realistic contours,
    \textcircled{\raisebox{-1pt} {$b$}}
    Blob Placement, where each generated blob is placed in a GT mask by a greedy algorithm that constrains blob density and distance to satisfy expert-driven criteria, and
    \textcircled{\raisebox{-1pt} {$c$}} 
    Image Generation via a style transfer Neural Network to generate visually realistic images from the computed GT. Experiments demonstrate that our method generates realistic images even starting from a single image, enabling training instance segmentation models with performance comparable to models trained on larger datasets.

    \section{Problem formulation}
    The Instance Segmentation problem is typically formulated as follows: given an histological image $\realimage \in \mathbb{R}^{\nrows \times \ncols \times 3}$, find all objects of a particular class (typically cells or part thereof), thus returning binary masks $\predictedmask \in \{0, 1\}^{\nrows \times \ncols \times \nblobs}$ where each channel contains a segmentation mask for one of $\nblobs$ objects within the image.
    Instance segmentation is typically solved by Convolutional Neural Networks (CNNs) trained on large datasets where each image $\realimage_\realdsindex$ is paired with its GT annotation $\realmask_\realdsindex$. 
    We address the problem of generating a training set composed of a large number ($N$) of synthetic images, $\genimageset$, starting from a small training set of real images $\realds$. In particular, each generated image $\genimage_\genindex$ is associated to a generated GT $\genmask_\genindex$, which can be used to train an instance segmentation network. Training set generation can be summarized as:
    \begin{equation*}
        \realds \rightarrow \gends, \quad \genIndex \gg \realdsIndex.
    \end{equation*}
 
    \section{Method}
    Fig.~\ref{fig:pipeline} and Algorithm \ref{alg:pipeline} describe the proposed generation pipeline that is composed of three parts. First, we leverage homotopy-based interpolation to generate blobs representing cells in a principled way; then, we generate GT segmentation masks by optimizing the positioning of blobs satisfying constraints on the blob distribution; lastly, we generate realistic images that match the generated GTs.
    
    \subsection{Blob Generation}
\begin{figure*}[t]
	  \centering
	  \subfloat[Homotopically equivalent blobs.]{\makebox[0.29\linewidth][c]{\label{fig:homotopy}\includegraphics[width=0.155\linewidth]{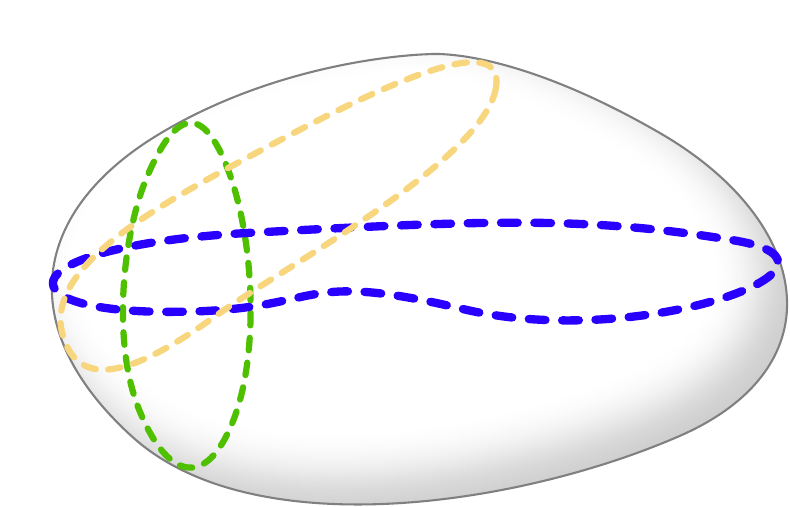}}}
	  \hfill
	  \subfloat[Interpolation lines between {\color{blue} $\blobone$} and {\color{OliveGreen} $\blobtwo$}]{\makebox[0.30\linewidth][c]{\label{fig:interpolation}\includegraphics[width=0.079\linewidth]{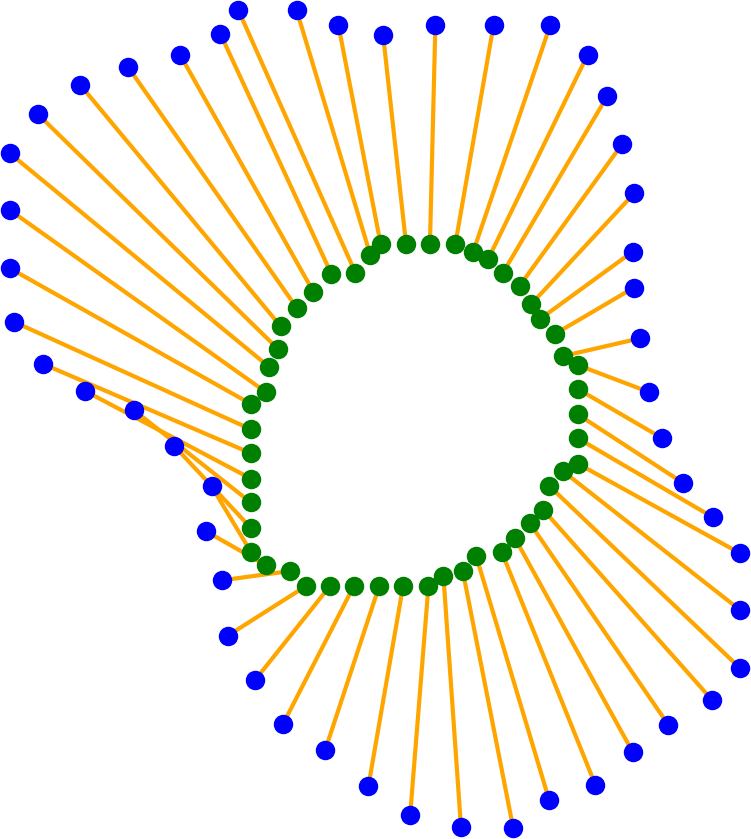}}}
	  \hfill
	  \subfloat[Interpolated blobs between {\color{blue} $\blobone$} and {\color{OliveGreen} $\blobtwo$}]{\makebox[0.4\linewidth][c]{\label{fig:interpolated}\includegraphics[width=0.35\linewidth]{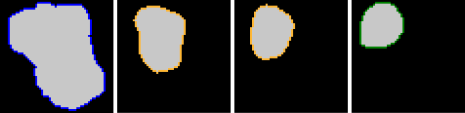}}}
	  \caption{Examples of interpolation between $\blobone$ (blue) and $\blobtwo$ (green). New blobs (in orange) are selected at equally spaced intervals along the interpolation lines and can be seen as different views of a 3D nucleus.}
	\end{figure*}

    The first step consists in generating a new set of $\genblobsIndex$ synthetic blobs $\genblobsset$  by interpolating pairs of cell masks randomly selected from real ones $\realblobsset$ (Fig.~\ref{fig:pipeline}, \textcircled{\raisebox{0.4pt} {$a$}}):
	$$
	\realblobsset \rightarrow \genblobsset, \quad \genblobsIndex \gg \realblobsIndex.
	$$
    The rationale behind our procedure is that blobs $\realblobsset$ in histopathology correspond to projections of 3D cells into an image plane. If we assume cells are 3D convex volumes, any pair of image projections from the same cell are \emph{homotopically equivalent}, thus projections over other planes can be obtained by continuously deforming one cell towards the other, as depicted in Fig.~\ref{fig:homotopy}. In non-convex cases, slicing a 3D cell may yield multiple connected components. Yet, homotopic equivalence applies locally to each projection with the same number of connected components. Since blobs in real images are projections of \emph{similar} 3D cells, we generate the contours of new realistic blobs by interpolating the contours of randomly selected pairs of blobs.
    Algorithm \ref{alg:interpolation} describes the procedure to generate $\genblobsIndex$ blobs starting from our set of real blobs $\realblobsset$ extracted from the training set. We first sample a random pair $\blobone$, $\blobtwo$ of real blobs, on which we identify a number $\perimeterpoints$ of equally spaced points along their contours $p_{k1}$ and $p_{k2}$. We then perform ICP registration \cite{besl_method_1992} to align and pair the contour points of the two blobs. Then, as shown in Fig.~\ref{fig:interpolation}, we interpolate between point pairs to generate the contour of the new blob:
    $$
      \widetilde{p}_\genblobsindex = \{\alpha p_{k1}^{(i)} + (1 - \alpha) p_{k2}^{(i)} \quad \forall i \in [1, E] \}, \,\,\alpha \in [0\,,\,1].
    $$
    Each generated blob $\genblobs$ is obtained by morphological area closing filling in the interpolated perimeters (Figure \ref{fig:interpolated}). The whole procedure runs in $\mathcal{O}(K \cdot L)$.

    \subsection{Blob Placement}
\begin{figure}[t]
        \centering
        \vspace{-10px}
        \subfloat[Prior map $\prior$\label{subfig:priormap}.]{\makebox[0.32\linewidth][c]{\label{fig:placement_prior}\includegraphics[width=0.155\textwidth]{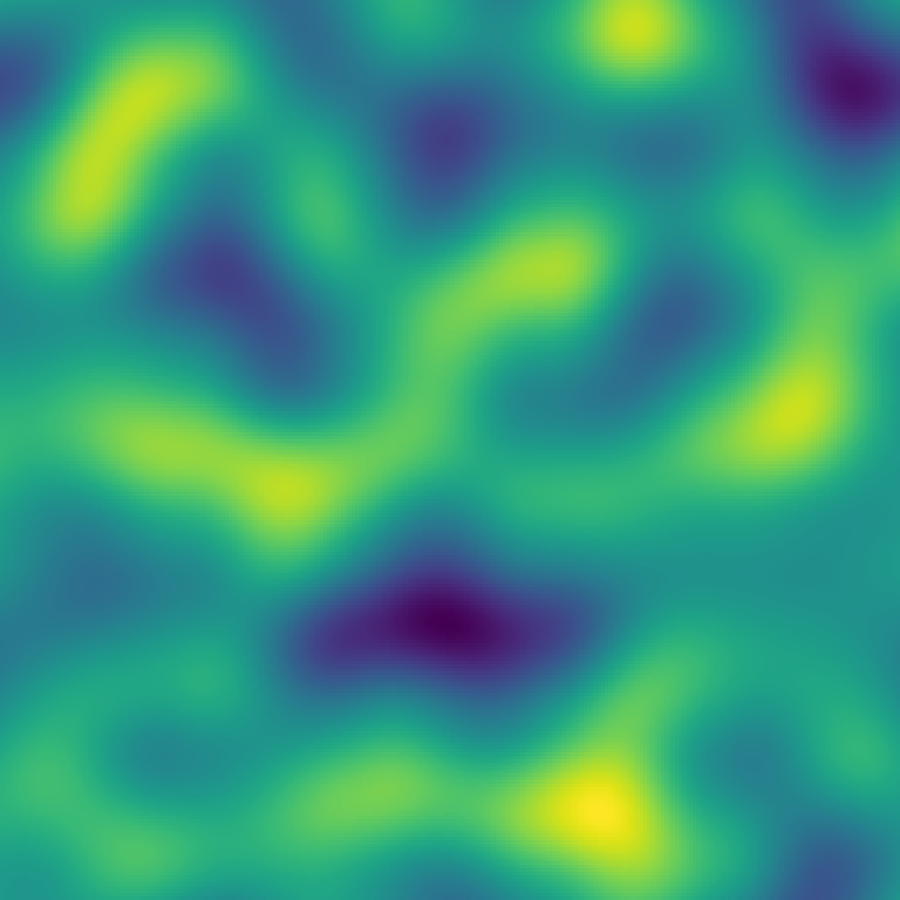}}}
        \hfill
        \subfloat[Random\label{subfig:randomplacement}.]{\makebox[0.32\linewidth][c]{\label{fig:randomplacement}\includegraphics[width=0.155\textwidth]{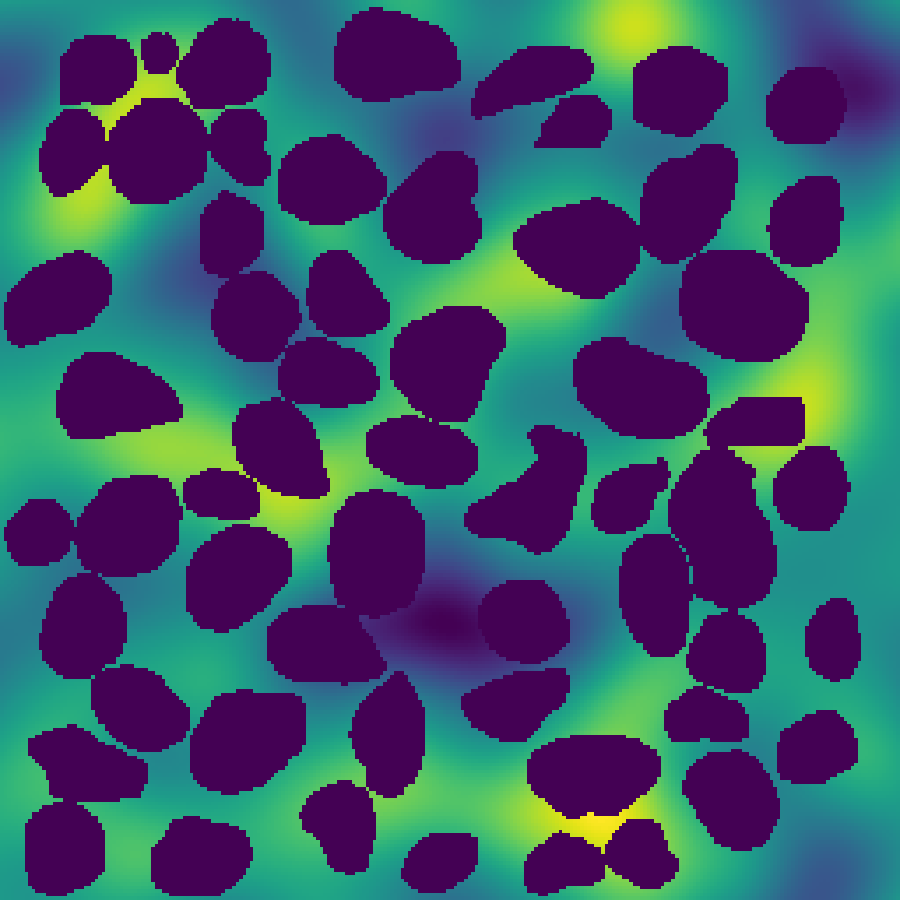}}}
        \hfill
        \subfloat[Greedy\label{subfig:greedyplacement}.]{\makebox[0.32\linewidth][c]{\label{fig:greedyplacement}\includegraphics[width=0.155\textwidth]{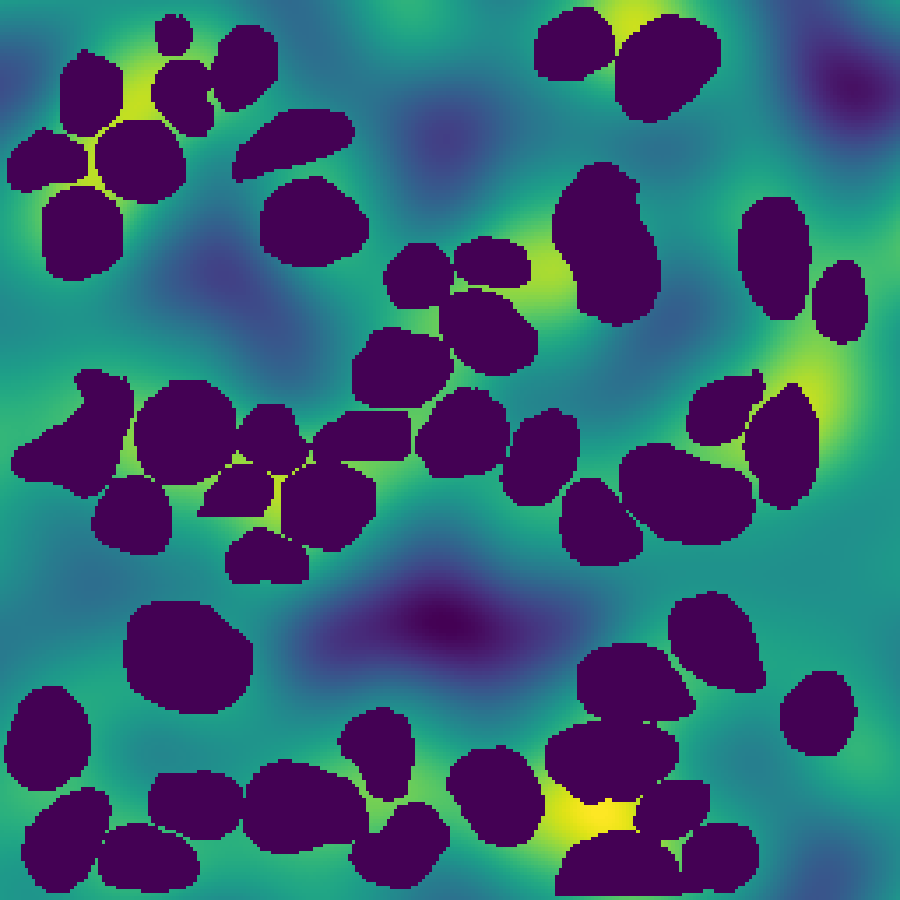}}}
        
        \caption{Examples of blob placement. Given a prior map $\prior$ (a), our greedy placement (c) respects much more closely the distribution with respect to a random weighted placement (b).}
		\label{fig:placement}
   \end{figure}

	% Intro
    After obtaining the set $\genblobsset$ of synthetic blobs, we define realistic image-level GTs $\genmaskset$ by a greedy blob placement procedure that enables controlling blob spacing and density (Figure~\ref{fig:pipeline}, \textcircled{\raisebox{-1pt} {$b$}}). To this end, we first define a set of 2D prior nuclei density distributions $\priorsset$, $\prior_\genindex \in [0, 1]^{\nrows \times \ncols}$.
    Each $\prior_\genindex$ is a heatmap  (Fig.~\ref{fig:placement_prior}), mimicking in each point $[i,j]$ the likelihood of a blob covering that region. Second, the univariate distribution $\priorspacing$ models the spacing between blobs in real images. Both $\prior_\genindex$ and $\priorspacing$ can be either estimated from the real available images or crafted by experts.

    The proposed blob placement procedure is described in  Algorithm \ref{alg:placement}. First, we initialize $\genmask$ as an empty binary mask and $\availablemask$ as an availability mask which is initially set to ones. Then, we iteratively place blobs by sampling locations in $\guidingmap_\genindex= \prior_\genindex \cdot \availablemask$, which is normalized to sum to 1 to mimic the distribution of currently available locations. More specifically we sample a point $[i,j]$ from  $\guidingmap_\genindex$ and a value from $\priorspacing$ representing the offset $z$. We then select the first generated blob in $\genblobsset$ that fits in at the sampled location. If no blob fits, the algorithm terminates. Otherwise, we update $\availablemask$ by setting to $0$ all the locations that are either covered by the blob or that are within a radius $z$ from $[i,j]$, enforcing a minimum spacing between blobs. We remove the placed blob from $\genblobsset$. At the end, the synthetic mask $\genmask$ is returned. As displayed in Fig.~\ref{fig:placement}, the proposed greedy procedure yields GTs that follow the prior more faithfully w.r.t. a random placement weighted according to $\prior$.

    Blob placement, although efficiently parallelizable, is the most computationally demanding component as it scales with $\mathcal{O}(\genIndex \cdot \frac{\nrows \cdot \ncols}{\blobarea} \cdot \genblobsIndex)$. The fractional term represents the number of blobs that can fit the image given its area and an average blob area $\blobarea$.

	\begin{algorithm}
		\begin{algorithmic}[1]
			\caption{Our generation pipeline}\label{alg:pipeline}
            \State \textbf{Input:} Small annotated training set $\realds$
            \State \textbf{Output:} Large synthetic training set $\gends$
			\State $\trainingset \gets \emptyset$  \Comment{Blobs interpolation}
			\State $\genblobsset \gets $ \Call{InterpolateBlobs}{$\realblobsset, \genblobsIndex, \perimeterpoints$}
            
			\For{$\genindex \in [1, \genIndex]$}
			\State $\prior, \priorspacing \gets$ \Call{GetPrior}{$\nrows, \ncols$} \Comment{Blobs placement}
			\State $\genmask_{\genindex} \gets$ \Call{GreedyPlacement}{$\prior, \genblobsset, \priorspacing$}
			\State $\refstyle \gets $ \Call{Sample}{$\realimageset$} \Comment{Style transfer}
			\State $\genimage_{\genindex} \gets$ \Call{AdaIN}{\textsc{Flatten}({$\genmask_{\genindex}$}), $\refstyle$}\label{algline:adain} 
			\State $\trainingset \gets \trainingset \cup \{(\genimage_\genindex, \genmask_\genindex)\}$
			\EndFor
		\end{algorithmic}
	\end{algorithm} 
 
	\begin{algorithm}
		\begin{algorithmic}[1]      
			\caption{Blob interpolation}\label{alg:interpolation}			
			\Procedure{InterpolateBlobs}{$\realblobsset, \genblobsIndex, \perimeterpoints$}
			\For{$\genblobsindex \in [1, \genblobsIndex]$}
			\State $\blobone, \blobtwo \gets$ \Call{Sample}{$\realblobsset, 2$}
			\State $p_1 \gets$ \Call{GetContourPoints}{$\blobone, \perimeterpoints$}
			\State $p_2 \gets$ \Call{GetContourPoints}{$\blobtwo, \perimeterpoints$}
			\State $p_1 \gets $ \Call{Registration}{$p_1, p_2$}
			\State $\alpha \gets $ \Call{Sample}{$[0, 1]$}
			\State $\widetilde{p} \gets$ \Call{Interpolate}{$p_1, p_2, \alpha$}
			\State $\genblobs_{\genblobsindex} \gets$ \Call{Closure}{$\widetilde{p}$}
			\EndFor
			\State \Return $\genblobsset$
			\EndProcedure
		\end{algorithmic}
	\end{algorithm}

	\begin{algorithm}
		\begin{algorithmic}[1]       
			\caption{Blob placement}\label{alg:placement}			
			\Procedure{GreedyPlacement}{$\prior_\genindex, \genblobsset, \priorspacing$}
			\State $\genmask \gets 0^{\nrows \times \ncols \times \nblobs}$
			\State $\availablemask \gets 1^{\nrows \times \ncols}$
			\While{$found$}
			\State $\guidingmap \gets $ \Call{Norm}{$\availablemask \cdot \prior_\genindex$}
   			\State $(y, x) \gets $ \Call{Sample}{$\guidingmap_\genindex$}
   			\State $z \gets $ \Call{Sample}{$\priorspacing$}
			\For{$b \in \genblobsset$}
			\State $found \gets$ \Call{CanHost}{$\guidingmap, b, y, x$}
			\If{$found$}
			\State $\genblobsset \gets \genblobsset \setminus \{b\}$
	           \State $\genmask \gets $ \Call{AddMask}{$\genmask, b, y, x$}
			\State $\availablemask \gets $ \Call{UpdateAvailable}{$\availablemask, b, z$}		
            \State \textbf{break}
			\EndIf
			\EndFor
			\EndWhile
			\State \Return $\genmask$
			\EndProcedure
		\end{algorithmic}
	\end{algorithm}

    \subsection{Image Generation}
    After obtaining the GT masks $\{\genmask_{\genindex}\}_\genIndex$, we generate the corresponding images by AdaIN \cite{huang_arbitrary_2017}, a style transfer framework (Figure~\ref{fig:pipeline}, \textcircled{\raisebox{0.4pt} {$c$}}) that can modify the style (i.e., the texture and appearance) of an image, while preserving its content. In our case, the content is the generated mask $\genmaskset$ while the style is represented by the collection of real images for training $\{\realimage_\realdsindex\}_\realdsIndex$. We therefore use AdaIN to create the texture characterizing cells over a mask $\genmask$, preserving the blob support resulting in pixel-perfect annotations for training segmentation networks. The content and style do not need to be paired, thus the collection of synthetic realistic images $\genimageset$ is obtained as follows:
    $$
    \genmaskset, \realimageset \rightarrow \gends.
    $$
    
\begin{figure}[t]
    	\centering
    	\subfloat[Generated mask $\genmask$.]{\makebox[0.31\linewidth][c]{\includegraphics[width=0.15\textwidth]{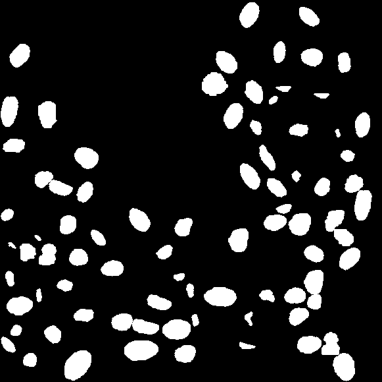}}}
    	\hfill
    	\subfloat[\raisebox{8pt}{\color{white}}Reference style $\refstyle$.]{\makebox[0.31\linewidth][c]{\includegraphics[width=0.15\textwidth]{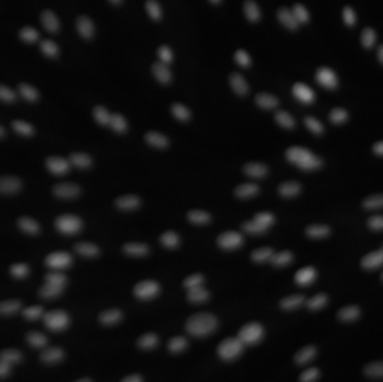}}}
    	\hfill
    	\subfloat[$\genimage$ generated from $\genmask$.]{\makebox[0.32\linewidth][c]{\includegraphics[width=0.15\textwidth]{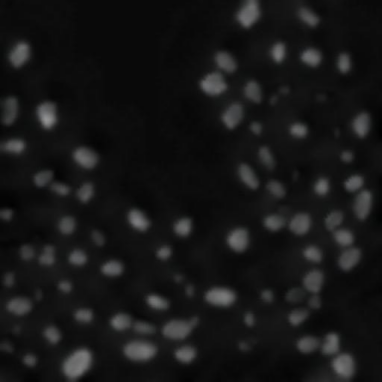}}}
        \caption{Example of style transfer.}
		\label{fig:transfer}
    \end{figure}

    In our pipeline (Algorithm \ref{alg:pipeline}, Line 9), we employ AdaIN by using as content the (flattened) generated mask $\genmask_\genindex$ and as reference style a patch $\refstyle$ from a real image $\realimage_\realdsindex$.  We train for 30000 epochs using as DA flips and affine transformations for both style and content, and photometric transformations for the style image only. The output of AdaIN is a realistic image $\genimage_\genindex$ where the displayed cells follow the input $\genmask_\genindex$. Pairs of generated image and mask can be used as annotated samples to augment the small annotated training set $\realds$. Fig.~\ref{fig:transfer} shows an example of this phase starting from a generated mask $\genmask$ and using as reference style an image $\refstyle$.

    \section{Experiments}
    We validate the effectiveness of our pipeline by training HoVerNet \cite{graham_hover-net_2019}, a state-of-the-art instance segmentation NN, on our generated dataset. We start by generating training sets from small subsets of
    fluorescence microscopy images from the Broad Institute Repository (BBBC) \cite{ljosa_annotated_2012}. We then assess the instance segmentation performance on real images from the NucleusSegData dataset \cite{gunesli_attentionboost_2020}.
    As a baseline, we compare the performance of the same network trained on the full BBBC dataset (\emph{Full dataset}) and on the few annotated images used for generation (\emph{Training real dataset}) leveraging standard augmentations. We adopt as metrics those used in HoVerNet, i.e., DICE, DICE2, AJI and AJI+.

    During the Blob Placement phase, we model $\prior_\genindex$ as realizations of Perlin noise \cite{perlin_improving_2002} as its complex 2D structures resemble a coarse view of the distribution of nuclei in histological images. We estimate the parameters of Perlin noise to maximize the similarity with blurred masks from training data. Then, we perform a preliminary analysis on the distributions of generated GTs to assess how close they are to real ones. In particular, we consider the distributions of area ($\mathcal{A}$) and aspect ratio ($ar$), reporting close results in both the median ($\mathcal{A}$: 155px vs. 153px, $ar$: $1.31$ vs. $1.41$) and in the Inter-Quartile Range ($\mathcal{A}$: 32px vs. 39px, $ar$: $0.21$ vs. $0.23$).
    
    For the image generation phase, we train AdaIN using a tiled version of real images. Then, for each generated mask $\genmask_\genindex$, we select as reference image the tile having the closest number of blobs to $\genmask_\genindex$. This choice improves the style transfer procedure, since AdaIN generates images having average value similar to the reference. 
    When the style has too many blobs, artefacts may appear in the image, and when it has too few, the nuclei may be fainter than in real images.

\begin{figure}
 	\centering
 	\resizebox{!}{0.5\linewidth}{
 		\begin{tikzpicture}
 			\begin{groupplot}[
 				group style={
 					group size=2 by 2,
 					horizontal sep=0.8cm,
 					vertical sep=0.8cm,
 					ylabels at=edge left, % Move ylabels to the left
 				},
 				width=0.75\textwidth,
 				height=0.4\textwidth,
 				y label style={font=\huge, at={(0.07,0.5)}},
 				xlabel style={font=\huge, at={(1.0,-0.01)}},
 				yticklabel style={anchor=west,xshift=1em, 
 					font=\Large},
 				xticklabel style={font=\Large},
 				tick pos=left,
 				legend style={at={(0.5,1.03)}, anchor=south, legend columns=-1},
 				]
 				
 				% Panel 1
 				\nextgroupplot[xlabel={}, ylabel={DICE}, ymax=0.99, xmax=952, xmin=1,ymax=1.02]
 				\addplot[color=blue, label={Real Dataset}] coordinates {(52,0.89) (321,0.89) (952,0.92)};
 				\addplot[color=green, label={Generated Dataset}] coordinates {(52,0.42) (321,0.89) (952,0.91)};
 				\draw[red, dashed] (axis cs: 0,0.94) -- (axis cs: 952,0.94);
 				\node[red, anchor=south east] at (axis cs: 900,0.98) {\huge $0.94$};
 				
 				% Panel 2
 				\nextgroupplot[xlabel={}, ylabel={DICE2}, ymin=0.46, xmax=952, xmin=1,ymax=0.732] % xtick=\empty
 				\addplot[color=blue, label={Real Dataset}] coordinates {(52,0.47) (321,0.70) (952,0.69)};
 				\addplot[color=green, label={Generated Dataset}] coordinates {(52,0.53) (321,0.58) (952,0.67)};
 				\draw[red, dashed] (axis cs: 0,0.70) -- (axis cs: 952,0.70);
 				\node[red, anchor=south east] at (axis cs: 900,0.715) {\huge $0.70$};
 				
 				% Panel 3
 				\nextgroupplot[xlabel={Number of Nuclei}, ylabel={AJI},ymax=0.90, xmax=952, xmin=1]
 				\addplot[color=blue, label={Real Dataset}] coordinates {(52,0.49) (321,0.73) (952,0.75)};
 				\addplot[color=green, label={Generated Dataset}] coordinates {(52,0.22) (321,0.58) (952,0.70)};
 				\draw[red, dashed] (axis cs: 0,0.81) -- (axis cs: 952,0.81);
 				\node[red, anchor=south east] at (axis cs: 900,0.85) {\huge $0.81$};
 				
 				% Panel 4
 				\nextgroupplot[xlabel={}, ylabel={AJI+}, ymax=0.90, xmax=952, xmin=1]
 				\addplot[color=blue, label={Real Dataset}] coordinates {(52,0.49) (321,0.74) (952,0.76)};
 				\addplot[color=green, label={Generated Dataset}] coordinates {(52,0.22) (321,0.58) (952,0.72)};
 				\draw[red, dashed] (axis cs: 0,0.81) -- (axis cs: 952,0.81);
 				\node[red, anchor=south east] at (axis cs: 900,0.85) {\huge $0.81$};
 				
 			\end{groupplot}
 			
 			% Define legend colors and styles
 			\definecolor{legendred}{RGB}{255,0,0}
 			\definecolor{legendgreen}{RGB}{0,128,0}
 			\definecolor{legendblue}{RGB}{0,0,255}

 			% Adjust the position of the legend
 			\begin{scope}[shift={(6.0,0.2)}]
 				% Draw white rectangle with black border
 				\draw[black, fill=white] (0,0) rectangle (5.5,2);
 				
 				% Draw legend lines and labels
 				\draw[legendred, dashed] (0.3,1.5) -- (0.8,1.5);
 				\draw[legendgreen] (0.3,1) -- (0.8,1);
 				\draw[legendblue] (0.3,0.5) -- (0.8,0.5);
 				
 				% Add legend labels
 				\node at (2.35,1.5) {\Large Full dataset};
 				\node at (2.4,1) {\Large Real dataset};
 				\node at (3.0,0.5) {\Large Generated dataset};
 			\end{scope}

 		\end{tikzpicture}
 	}
 	\caption{Our results by metric per number of nuclei instances in the real training set.}
 	\label{fig:results}
 \end{figure}
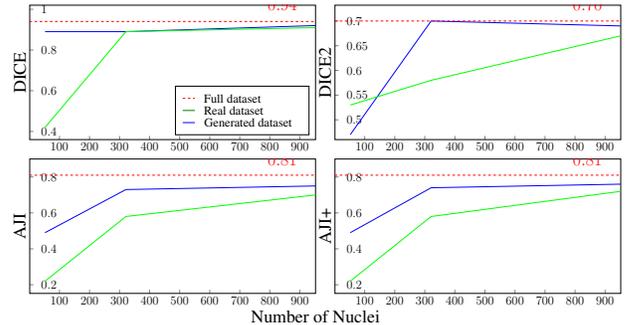

    Results shown in Table \ref{tab:results} and in Figure \ref{fig:results} indicate that HoVerNet trained on the full dataset ($815$ annotated images, $\approx 50\,000$ blobs) achieves an impressive $0.94$ DICE score and $0.81$ AJI score. These values need to be considered as an ideal standard and are displayed with a dashed line in Figure \ref{fig:results}. When trained from a generated dataset starting from only two real images ($52$ blobs), HoVerNet achieves $0.89$ DICE score and $0.49$ AJI score, compared to $0.42$ and $0.22$ when the same architecture is trained only on the same two real images. When increasing the number of real images this gap decreases, and the advantages of image generation are lost starting from $10$ real images ($\approx 1\,000$ blobs).
    Nonetheless, these results show that our pipeline enables to train models effectively even in very low data regimes.
	
	\begin{table}[t]
		\centering 
		\caption{Results on our test set.}
		\begin{tabular}{| c | c | c | c | c | c |}
			\hline
			\textbf{$\realdsIndex$} & \textbf{$\realblobsIndex$} & \textbf{DICE} & \textbf{DICE2} & \textbf{AJI} & \textbf{AJI+} \\ \hline\hline
			\multicolumn{6}{| c |}{Full dataset}                                                                    \\ \hline
			815                     & 46016                      & 0.94          & 0.70           & 0.81         & 0.81          \\ \hline\hline
			\multicolumn{6}{| c |}{Training on real images}                                                   \\ \hline
			2                       & 52                         & 0.42          & 0.53           & 0.22         & 0.22          \\
			5                       & 321                        & 0.89          & 0.58           & 0.58         & 0.58          \\
			10                      & 952                        & 0.91          & 0.67           & 0.70         & 0.72          \\ \hline
			\multicolumn{6}{| c |}{Training on our generated images}                                         \\ \hline
			2                       & 52                         & 0.89          & 0.48           & 0.49         & 0.49          \\
			5                       & 321                        & 0.89          & 0.70           & 0.73         & 0.74          \\
			10                      & 952                        & 0.92          & 0.69           & 0.75         & 0.76          \\ \hline
		\end{tabular}
		\label{tab:results}
	\end{table}

    \section{Conclusions and future work}
    We presented a novel pipeline for generating synthetic and realistic images, paired with their annotation for training instance segmentation networks. 
    Our approach generates realistic blobs by interpolation and enables experts to control the density and spacing of blob placement in GTs. 
    We demonstrated that our data generation pipeline helps in training a NN in low-data regimes. 

    Our future work consists in assessing the potential of our method to counteract domain shift with very scarce annotations. We will also explore GANs conditioned on mask geometric properties to promote diversity in the nuclei shapes. %We also would like to study the impact of different prior probabilities in the placement phase while using more sophisticated optimization techniques able to account for small overlaps as well.
    Additionally, we will analyze how different priors can influence the placement phase, and employ advanced optimization techniques capable of accounting for small overlaps as well.
    
    \section{Compliance with Ethical Standards}
    This research study was conducted retrospectively using data made available in open access. Ethical approval was not required as confirmed by the license attached with the open access data.
	
    \section{Acknowledgments}
    This paper is supported by FAIR (Future Artificial Intelligence Research) project, funded by the NextGenerationEU program within the PNRR-PE-AI scheme (M4C2, Investment 1.3, Line on Artificial Intelligence).

    We also would like to thank Ikonisys, Inc. for sponsoring the PhD grant of Roberto Basla and for an insightful discussion.
 
	\bibliographystyle{IEEEbib}
	\bibliography{strings,refs}	

\begin{thebibliography}{10}

\bibitem{shorten_survey_2019}
Connor Shorten and Taghi~M. Khoshgoftaar,
\newblock ``A survey on {Image} {Data} {Augmentation} for {Deep} {Learning},''
\newblock {\em Journal of Big Data}, vol. 6, no. 1, pp. 60, Dec. 2019.

\bibitem{mahmood_deep_2018}
Faisal Mahmood, Daniel Borders, Richard Chen, Gregory~N. McKay, Kevan~J.
  Salimian, Alexander Baras, and Nicholas~J. Durr,
\newblock ``Deep {Adversarial} {Training} for {Multi}-{Organ} {Nuclei}
  {Segmentation} in {Histopathology} {Images},'' Oct. 2018,
\newblock arXiv:1810.00236 [cs].

\bibitem{li_high_2022}
Wenyuan Li, Jiayun Li, Jennifer Polson, Zichen Wang, William Speier, and Corey
  Arnold,
\newblock ``High resolution histopathology image generation and segmentation
  through adversarial training,''
\newblock {\em Medical Image Analysis}, vol. 75, pp. 102251, Jan. 2022.

\bibitem{goodfellow_generative_2014}
Ian~J. Goodfellow, Jean Pouget-Abadie, Mehdi Mirza, Bing Xu, David
  Warde-Farley, Sherjil Ozair, Aaron Courville, and Yoshua Bengio,
\newblock ``Generative {Adversarial} {Networks},'' June 2014,
\newblock arXiv:1406.2661 [cs, stat].

\bibitem{hou_robust_2019}
Le~Hou, Ayush Agarwal, Dimitris Samaras, Tahsin~M. Kurc, Rajarsi~R. Gupta, and
  Joel~H. Saltz,
\newblock ``Robust {Histopathology} {Image} {Analysis}: {To} {Label} or to
  {Synthesize}?,''
\newblock in {\em 2019 {IEEE}/{CVF} {Conference} on {Computer} {Vision} and
  {Pattern} {Recognition} ({CVPR})}, Long Beach, CA, USA, June 2019, pp.
  8525--8534, IEEE.

\bibitem{cheng_deep_2021}
Jijun Cheng, Zimin Wang, Zhenbing Liu, Zhengyun Feng, Huadeng Wang, and Xipeng
  Pan,
\newblock ``Deep {Adversarial} {Image} {Synthesis} for {Nuclei} {Segmentation}
  of {Histopathology} {Image},''
\newblock in {\em 2021 2nd {Asia} {Conference} on {Computers} and
  {Communications} ({ACCC})}, Singapore, Sept. 2021, pp. 63--68, IEEE.

\bibitem{kromp_evaluation_2021}
Florian Kromp, Lukas Fischer, Eva Bozsaky, Inge~M. Ambros, Wolfgang Dorr, Klaus
  Beiske, Peter~F. Ambros, Allan Hanbury, and Sabine Taschner-Mandl,
\newblock ``Evaluation of {Deep} {Learning} {Architectures} for {Complex}
  {Immunofluorescence} {Nuclear} {Image} {Segmentation},''
\newblock {\em IEEE Transactions on Medical Imaging}, vol. 40, no. 7, pp.
  1934--1949, July 2021.

\bibitem{hou_unsupervised_2017}
Le~Hou, Ayush Agarwal, Dimitris Samaras, Tahsin~M. Kurc, Rajarsi~R. Gupta, and
  Joel~H. Saltz,
\newblock ``Unsupervised {Histopathology} {Image} {Synthesis},'' Dec. 2017,
\newblock arXiv:1712.05021 [cs].

\bibitem{besl_method_1992}
P.J. Besl and Neil~D. McKay,
\newblock ``A method for registration of 3-{D} shapes,''
\newblock {\em IEEE Transactions on Pattern Analysis and Machine Intelligence},
  vol. 14, no. 2, pp. 239--256, Feb. 1992.

\bibitem{huang_arbitrary_2017}
Xun Huang and Serge Belongie,
\newblock ``Arbitrary {Style} {Transfer} in {Real}-time with {Adaptive}
  {Instance} {Normalization},'' July 2017,
\newblock arXiv:1703.06868 [cs].

\bibitem{graham_hover-net_2019}
Simon Graham, Quoc~Dang Vu, Shan E~Ahmed Raza, Ayesha Azam, Yee~Wah Tsang,
  Jin~Tae Kwak, and Nasir Rajpoot,
\newblock ``Hover-{Net}: {Simultaneous} segmentation and classification of
  nuclei in multi-tissue histology images,''
\newblock {\em Medical Image Analysis}, vol. 58, pp. 101563, Dec. 2019.

\bibitem{ljosa_annotated_2012}
Vebjorn Ljosa, Katherine~L Sokolnicki, and Anne~E Carpenter,
\newblock ``Annotated high-throughput microscopy image sets for validation,''
\newblock {\em Nature Methods}, vol. 9, no. 7, pp. 637--637, July 2012.

\bibitem{gunesli_attentionboost_2020}
Gozde~Nur Gunesli, Cenk Sokmensuer, and Cigdem Gunduz-Demir,
\newblock ``\textit{{AttentionBoost}} : {Learning} {What} to {Attend} for
  {Gland} {Segmentation} in {Histopathological} {Images} by {Boosting} {Fully}
  {Convolutional} {Networks},''
\newblock {\em IEEE Transactions on Medical Imaging}, vol. 39, no. 12, pp.
  4262--4273, Dec. 2020.

\bibitem{perlin_improving_2002}
Ken Perlin,
\newblock ``Improving noise,''
\newblock in {\em Proceedings of the 29th annual conference on {Computer}
  graphics and interactive techniques}, San Antonio Texas, July 2002, pp.
  681--682, ACM.

\end{thebibliography}
	
\end{document}